\documentclass[11pt]{article}
\usepackage{amsmath,amssymb,color,graphics,epsfig}

\usepackage{amsfonts}

\newcommand{\be}{\begin{equation}}
\newcommand{\ee}{\end{equation}}
\newcommand{\bea}{\setlength\arraycolsep{2pt} \begin{eqnarray}}
\newcommand{\eea}{\end{eqnarray}}
\newcommand{\nn}{\nonumber}

\newcommand{\mm}{\mathrm}

\def\fft#1#2{{\frac{#1}{#2}}}

\def\0{{\sst{(0)}}}
\def\1{{\sst{(1)}}}
\def\2{{\sst{(2)}}}
\def\3{{\sst{(3)}}}
\def\4{{\sst{(4)}}}
\def\5{{\sst{(5)}}}
\def\6{{\sst{(6)}}}
\def\7{{\sst{(7)}}}
\def\8{{\sst{(8)}}}
\def\sst#1{{\scriptscriptstyle #1}}

\begin{document}

\begin{flushright}
\end{flushright}

\vspace{25pt}
\begin{center}
{\large {\bf Critical phenomenon of quantum BTZ black holes}}

\vspace{10pt}
 Hong-Ming Cui$^{1\dagger}$ and Zhong-Ying Fan$^{1\dagger}$

\vspace{10pt}
$^{1\dagger}${ Department of Astrophysics, School of Physics and Material Science, \\
 Guangzhou University, Guangzhou 510006, P.R. China }\\

\vspace{40pt}

\underline{ABSTRACT}
\end{center}
We extend the thermodynamics of quantum BTZ black holes by treating the quantum backreaction strength parameter $\nu$ as a thermodynamic variable. We find various novel features. The critical point appears at $\nu_c=1$ and a first order transition occurs either below the critical temperature for $\nu<\nu_c$ or above the critical temperature for $\nu>\nu_c$. By solving the coexistence curve analytically, we analyze the phase structures and clarify an unexpected discontinuity around the critical point. The critical exponents are significantly different from the mean field theory results and violate one of the scaling laws. We present an intepretation for this by using a universal three scale factor hypothesis for critical behavior of thermodynamic potential. Finally, we prove that given an arbitrarily small angular momenta, only one stable black hole phase can exist and hence no transition will occur.

\vfill {\footnotesize  Email: fanzhy@gzhu.edu.cn\,.}

\thispagestyle{empty}

\pagebreak

\tableofcontents
\addtocontents{toc}{\protect\setcounter{tocdepth}{2}}




\section{Introduction}
The extended thermodynamics of Anti-de-Sitter (AdS) black holes was developed about a decade ago. In the pioneer work \cite{Kubiznak:2012wp},  the cosmological constant was identified to a thermodynamic pressure and its conjugate as a thermodynamic volume. Appearance of these new notions has facilitated numerous developments in black hole physics, from macroscopically to microscopically. For example, using these notions, the reverse isoperimetric inequality proposes a new upper bound for black hole entropy \cite{Cvetic:2010jb} (generalization to rotating black holes was given in \cite{Amo:2023bbo}) whereas the Ruppeiner geometry associated to thermodynamic fluctuations of these variables illustrates the microstructure interactions of AdS black holes partly, independent of any microscopic foundation \cite{Wei:2015iwa,Wei:2019uqg}. Besides, the spirit of extended thermodynamics is far beyond a specific coupling constant and actually is valid to all coupling constants. By varying the Newton's gravitational constant, a new pair of thermodynamic variables: the central charge in the boundary CFT and its conjugate, was recently introduced for charged AdS black holes \cite{Cong:2021fnf,Cong:2021jgb}.

However, most studies in this field are focus on classical AdS black holes, corresponding to holographic duals of boundary theories in the large N limit. In this case, a black hole seems like a Van-der Waals fluid and critical behaviors of various physical quantities are captured by the mean field theory. This is simple but is less interesting in physics. Recently, to go beyond the large N limit, the extended thermodynamics of quantum corrected black holes has been studied in literature \cite{Frassino:2023wpc,HosseiniMansoori:2024bfi,Hu:2024ldp}. It is established that these black holes indeed look like some quantum fluids, which invalid the mean field theory.

Inspired by these works, we are interested in studying critical phenomenon of the (rotating) quantum BTZ black holes (qBTZ), which capture full quantum backreactions of conformal fields \cite{Emparan2020}. We extend the thermodynamics by treating the quantum backreaction strength parameter $\nu$ as a thermodynamic variable. We find various novel features, which are not discovered in a previous work \cite{HosseiniMansoori:2024bfi}. The critical point appears at $\nu_c=1$ and a first order transition occurs either below the critical temperature for $\nu<\nu_c$ or above the critical temperature for $\nu>\nu_c$. As a comparison, for a Van-der Waals like fluid, the transition terminates above the critical temperature. We solve the coexistence curve analytically and analyze the phase structures carefully. In particular, we clarify emergence of an unexpected discontinuity around the critical point. We compute various critical exponents and find that they are significantly different from the mean field theory results. In particular, they violate one of the scaling laws. We try to interpret this by developing a universal three scale factor hypothesis for critical behavior of thermodynamic potential. If our interpretation is correct, there will exist a more general constraint on the critical exponents (see (\ref{generalconstaint}) and discussions above). Finally, we prove that given an arbitrarily small angular momenta, only one stable black hole phase can exist and hence no transition will occur.

The remainder of this paper is organized as follows. In section 2, we briefly review the thermodynamics of generally rotating quantum BTZ black holes and extend it by taking the quantum backreaction strength parameter $\nu$ as a thermodynamic variable. In section 3, we study critical phenomenon associated to $\nu$ and its conjugate chemical potential $U$ in details. We report various novel features that are mentioned before. In section 4, we compute various critical exponents and establish that they violate one of the scaling law. To interprete this, we develop a three scale factor hypothesis for thermodynamic potential.  In section 5, we explore the role of a small angular momenta on the $U-\nu$ criticality.

\section{Extended thermodynamics of rotating qBTZ black hole}
The quantum corrected BTZ black hole was holographically interpreted as a
black hole localized on a Karch-Randall brane in the classical AdS$_4$ C-metric \cite{Emparan2020}. The position of the brane is described by a length parameter $\ell$, which is inversely proportional the brane tension. This parameter characterizes holographically the strength of quantum backreactions of conformal fields in AdS$_3$. The ordinary thermodynamics of the solution has been studied in \cite{Emparan2020}. The mass $M$, the temperature $T$, the entropy $S$, the angular velocity $\Omega$ and the angular momenta $J$ are given by
\bea
M&=&\fft{1}{2G_3}\fft{\sqrt{1+\nu^2}(1-z^3\nu)\left(z^2(1+z\nu)+\alpha\left(1+4z^2+4z^3\nu(1+\alpha^2)-z^4(1+4\alpha^2)\right)\right)}{\left(1+3z^2+2z^3\nu-\alpha^2(1-4z^3\nu+3z^4)\right)^2}\,,\nn\\
T&=&\fft{1}{2\pi l_3}\fft{\left(z^2(1+z\nu)-\alpha^2(1-2z^3\nu+z^4)\right)\left(2+3z\nu(1+\alpha^2)-4\alpha^2z^2+z^3\nu+\alpha^2 z^5\nu\right)}{z(1+z\nu)\left(1+\alpha(1-z^2)\right)\left(1+3z^2+2z^3\nu-\alpha^2(1-4z^3\nu+3z^4)\right)} \,,\nn\\
S&=&\fft{\pi l_3}{G_3}\fft{z\sqrt{1+\nu^2}\left(1+\alpha^2(1-z^2)\right)}{\left(1+3z^2+2z^3\nu-\alpha^2(1-4z^3\nu+3z^4)\right)} \,,\nn\\
\Omega&=&\fft{1}{l_3}\fft{\alpha(1+z^2)\sqrt{(1-z^3\nu)\left(1+z\nu-\alpha^2 z(z-\nu)\right)}}{z(1+z\nu)\left(1+\alpha^2(1-z^2)\right)} \,,\nn\\
J&=&\fft{l_3}{G_3}\fft{\alpha z\sqrt{1+\nu^2}(1+z^2)\left(1+\alpha^2(1-z^2)\right)\sqrt{(1-z^3\nu)\left(1+z\nu-\alpha^2z(z-\nu)\right)}}{\left(1+3z^2+2z^3\nu-\alpha^2(1-4z^3\nu+3z^4)\right)^2}\,,
\label{rotating quantities}
\eea
where $\alpha$ is the rotation parameter, $z$ is related to the event horizon radius $r_+$ and the strength of quantum backreactions is normalized by the AdS$_3$ radius $\ell_3$
\bea
z\equiv\fft{l_3}{r_+x_1}\,,\quad\quad \nu\equiv\fft{l}{l_3}\,,
\eea
where $x_1$ is a parameter describing the truncated AdS$_4$ C-metric having a finite black hole in the bulk. Notice that $0
<\nu<+\infty$ whereas the black hole size $z$ is strongly constrained by the rotation (in the static limit $z$ runs in the region $(0\,,+\infty)$  ). The entropy of the qBTZ black hole was holographically interpreted as the quantum corrected generalized entropy $S_{\mm{gen}}$, which includes the contributions of quantum entanglement of conformal fields outside the event horizon. It was established in \cite{Emparan2020} that the first law $dM=T dS+\Omega dJ$ holds. 

In this work, we would like to extend the thermodynamics of qBTZ black holes by varying the strength of quantum backreactions and explore the critical phenomenon accordingly. It is known that in the absence of quantum backreactions $\nu\rightarrow 0$, the classical BTZ black holes do not exhibit any phase transition. It will be very interesting to see whether or how critical phenomenon emerges in the presence of quantum backreactions. For this purpose, we will directly take the quantum backreaction parameter $\nu$ as a thermodynamic variable and fix both the AdS$_3$ radius and the Newton's gravitational constant $G_3$ throughout this work. In this situation, the chemical potential $U$ conjugate to $\nu$ can be evaluated to be
\bea
U&=&\Big[ \alpha^2(2\nu^2+1)z^5-\nu^3(2\alpha^2+1)z^4-\big( 3\nu^2+2\alpha^2(\nu^2+1)+1\big)z^3 +2\nu \alpha^2 z^2   \\
&&+(\alpha^2-\nu^2+1)z-\nu \Big] 
\times  \fft{\Big(-\alpha^2z^4+\nu(2\alpha^2+1)z^3+z^2-\alpha^2\Big)}{2G_3(1+\nu z)\left(3\alpha^2z^4-2\nu(2\alpha^2+1)z^3-3z^2+\alpha^2-1\right)^2\sqrt{1+\nu^2}}\,.\nn
\eea
It follows that the extended first law
\bea
dM=TdS+\Omega dJ+Ud\nu\,,
\eea
holds. Despite that the thermodynamic pressure $P$ and the central charge $C$ are not written explicitly in our extended first law, they will actually vary as $\nu$ varies \cite{HosseiniMansoori:2024bfi}. However, these effects are attributed to the conjugate chemical potential $U$ so we will not discuss them any more. 

It should be emphasized that in the presence of rotation, the parameter region in which a qBTZ black hole exists is complicated and not fully explored \cite{Emparan2020}. However, for our purpose we will focus on studying the correction effects from a small angular momenta. In this case, the black hole size is simply constrained to be $z\leq \nu^{-1/3}$. The leading effect of a small $J$ to the critical phenomenon will be investigated in section 5 in details.


\section{Critical phenomenon of static qBTZ black holes}

\begin{figure}
  \centering
  \includegraphics[width=270pt]{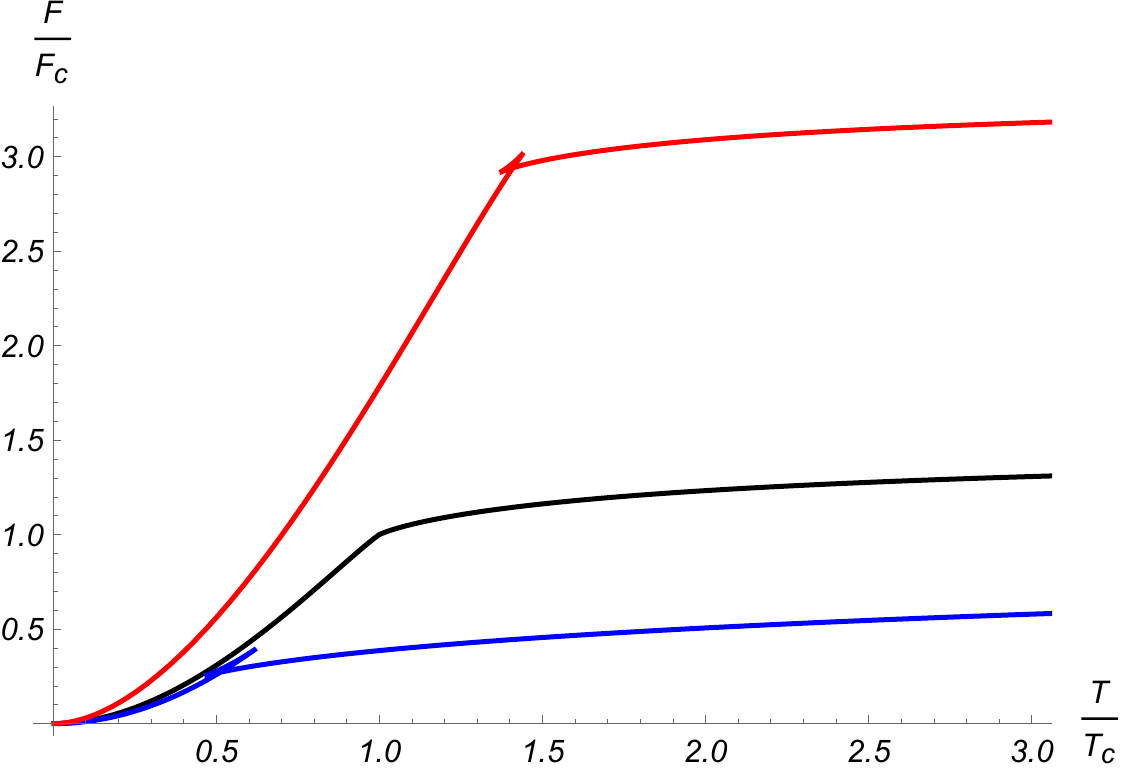}
  \caption{The behavior of free energy for a static qBTZ black hole. $\nu=\nu_c=1$ (black), $\nu=0.1$ (blue) and $\nu=3$ (red).}
  \label{t-f t-u}
\end{figure}

Consider the static limit at first. The various thermodynamic parameters simplify to 
\bea
M&=&\fft{\sqrt{1+\nu^2}}{2G_3}\fft{z^2(1-z^3\nu)(1+z\nu)}{(1+3z^2+2z^3\nu)^2}\,,\nn\\
T&=&\fft{z(2+3\nu z+\nu z^3)}{2\pi l_3(1+3z^2+2\nu z^3)} \,,\nn\\
S&=&\fft{\pi l_3}{G_3}\fft{z\sqrt{1+\nu^2}}{1+3z^2+2\nu z^3}  \,,\nn\\
U&=&-\fft{z^2\left(\nu+z^4\nu^3+z(\nu^2-1)+z^3(3\nu^2+1)\right)}{2G_3\sqrt{1+\nu^2}(1+3z^2+2z^3\nu)^2}  \,.
\eea
 Using the inflection point condition $\fft{\partial T}{\partial S}=\fft{\partial^2T}{\partial S^2}=0$ for fixed $\nu$, we deduce
\bea
\Big(\fft{\partial T}{\partial S}\Big)_\nu&=&\fft{G_3}{\pi^2 l_3^2}\fft{(1-z^3\nu)(z^3\nu+3z^2-3z\nu-1)}{\sqrt{1+\nu^2}(4z^3\nu+3z^2-1)} \,,\nn\\
\Big(\fft{\partial^2T}{\partial S^2}\Big)_\nu&=&\fft{G_3^2}{\pi^3 l_3^3}\fft{3\nu(z^2-1)(2z^3\nu+3z^2+1)^4}{(1+\nu^2)(4z^3\nu+3z^2-1)}\,.
\eea
Clearly the critical point appears at $z_c=1\,,\nu_c=1$. The other critical parameters are given by
\bea
T_c=\fft{1}{2\pi l_3}\,,\quad S_c=\fft{\pi l_3}{3\sqrt{2}G_3}\,,\quad U_c=-\fft{1}{12\sqrt{2}G_3}\,.
\eea
To explore whether a first order transition occurs when $\nu\neq \nu_c$, we depict the behavior of free energy as a function of the temperature $T$ in Fig. \ref{t-f t-u}. Surprisingly, we find that for both $\nu<\nu_c$ and $\nu>\nu_c$, the characteristic swallow tail behavior emerges, implying that the small-large black hole transition exists in both cases.  

\subsection{Coexistence curve and the phase diagram}
\begin{figure}
  \centering
  \includegraphics[width=210pt]{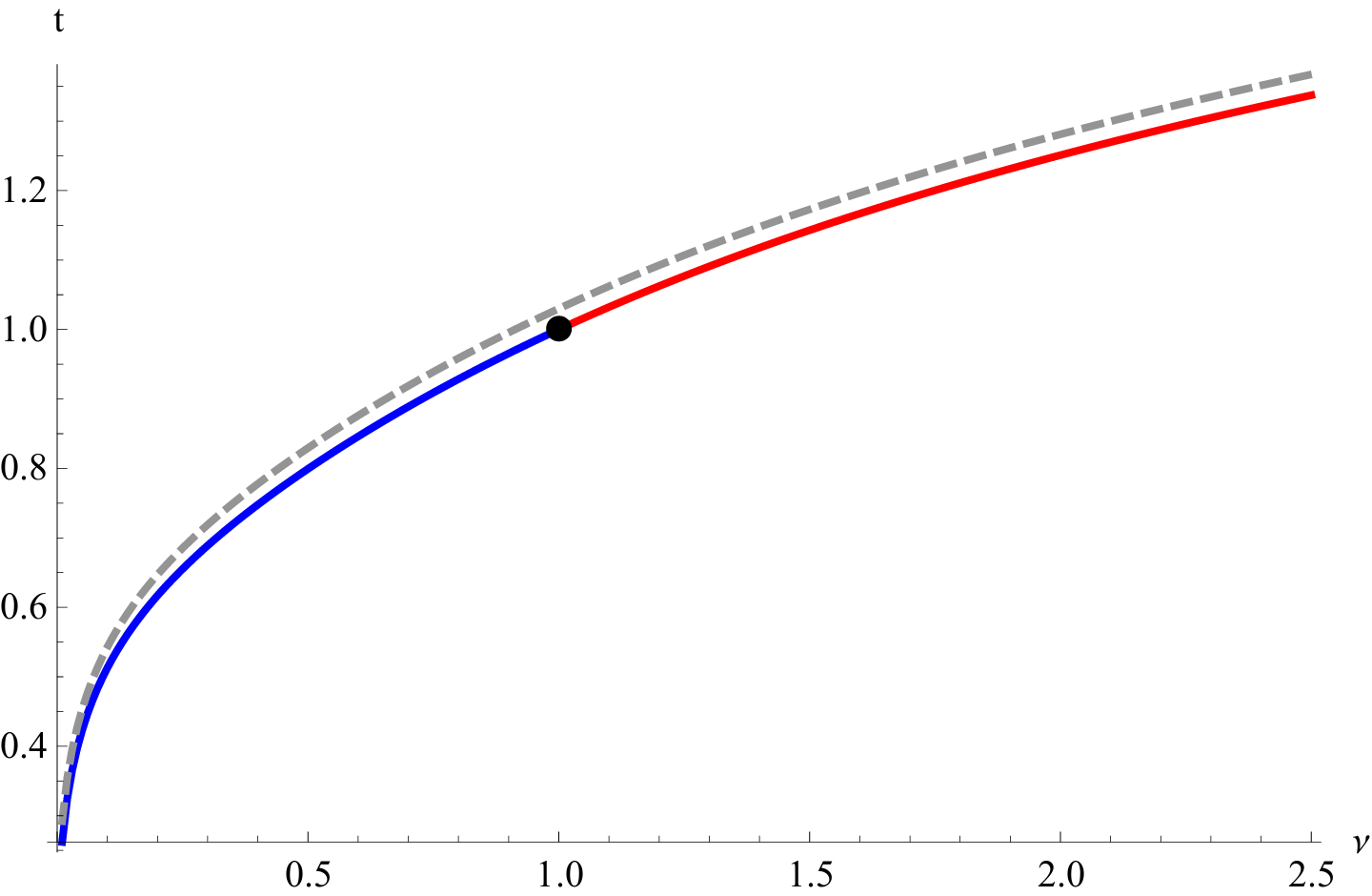}
  \includegraphics[width=210pt]{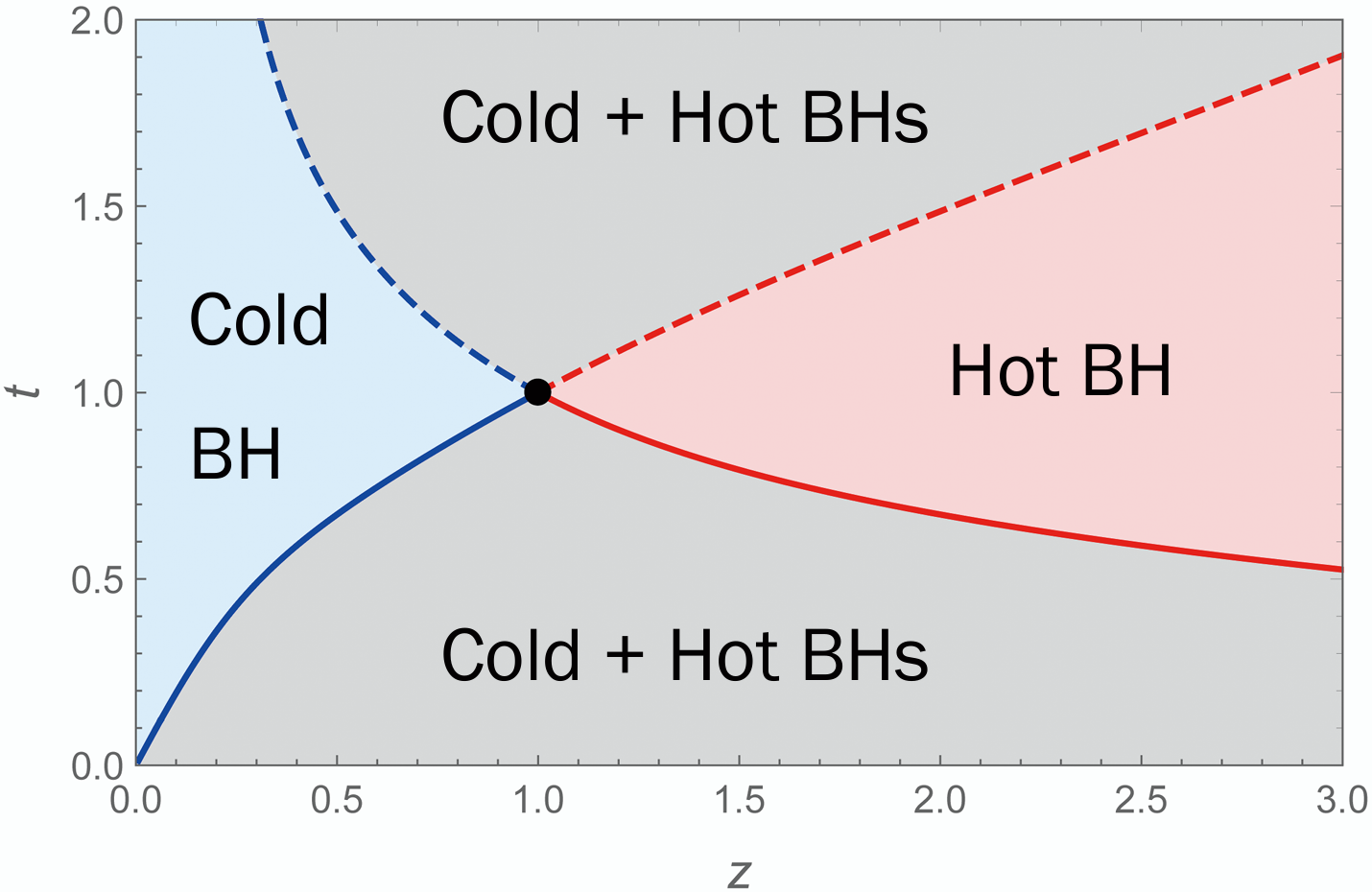}
  \caption{Left: the coexistence curve on the $t-\nu$ plane. The blue (red) curve corresponds to $\nu\le1 (\nu\ge1)$. The dashed line stands for the numerical solution, which has been moved along the vertical axis slightly. Right: the phase structure of the static qBTZ black hole on the $t-z$ plane. Solid curves describe the $\nu<1$ case, whereas the dashed ones describe the $\nu>1$ case, respectively. The three phases are represented by the shaded regions with different color.}
\label{ts(nu)}
\end{figure}
To confirm our new findings, we move to compute the coexistence curve for both $\nu<\nu_c$ and $\nu>\nu_c$ region. For simplicity, we work with the normalized temperature and free energy
 $t=T/T_c\,,f=F/F_c$:
\bea\label{t,s,u}
t&=&\fft{z(2+3z\nu+z^3\nu)}{1+3z^2+2z^3\nu}\,,\nn\\
f&=&\fft{3\sqrt{2}\sqrt{1+\nu^2}z^2\Big(1+z\nu\big(2+z^2(2+z\nu) \big) \Big)}{\Big(1+z^2(3+2z\nu) \Big)^2}\,.
\eea 
It turns out that we can obtain the coexistence curve analytically by solving the equations $f(z_s)=f(z_l),\, t(z_s)=t(z_l)$ carefully, where $z_s(z_l)$ stands for the small (large) black hole size, respectively (here by ``small'', we mean $z_s>1$ and by ``large'' $z_l<1$, compared to the critical point). The non-trivial solutions to these equations satisfy
\bea\label{small solution}
z_s\,z_l=1\,, \quad\quad 8\nu=(z_*+1/z_*)^3\pm\sqrt{(z_*+1/z_*)^6-64}\,,
\eea
where $z_*=z_s\,(\mm{or}\,z_l)$  and the ``$\pm$'' sign corresponds to $\nu\geq 1$ and $\nu \leq 1$, respectively. However, for both cases, one has (the unphysical solution has been dropped)
\be\label{zslnu} z_*(\nu)=\fft12\left( \varphi(\nu)\pm\sqrt{\varphi(\nu)^2-4} \, \right)\,,\qquad \varphi^3=4(\nu+1/\nu) \,,\ee
where the sign ``$\pm$'' corresponds to the small and the large black hole respectively. Substituting the result into the equation of state, we derive the coexistence curve on the $t-\nu$ plane
\be t_*(\nu)=\fft{z_*(\nu)\Big(2+3z_*(\nu)\nu+z_*^3(\nu)\nu \Big)}{1+3z_*^2(\nu)+2z_*^3(\nu)\nu } \,.\ee
The result is established as the solid line in the left panel of Fig. \ref{ts(nu)} and is perfectly matched with the numerical result (shown in dashed line). It is interesting to observe that for $\nu\geq 1$, the small-large black hole transition occurs for $T\geq T_c$ whereas for $\nu \leq 1$, the transition occurs for $T\leq T_c$, despite that the coexistence curve (and its slope) is smooth across the critical point. Critical phenomenon of this type was found previously for the quantum anomaly corrected black holes \cite{Hu:2024ldp} but was unfortunately ignored in \cite{HosseiniMansoori:2024bfi}.  

Furthermore, to study the phase structure of qBTZ black holes, we deduce the coexistence curve on the $t-z$ plane by using (\ref{small solution})
\bea\label{t*(z)}
t_*
=\fft{z_*^2\left(3+z_*^2\right)\left(\sqrt{(z_*+z_*^{-1})^6-64}\pm (z_{*}+z_*^{-1})^3\right)\pm 16z_*}{2z_*^3\left(\sqrt{(z_*+z_*^{-1})^6-64}\pm (z_*+z_*^{-1})^3\right)\pm 8\left(1+3z_*^2\right)}
\,,
\eea
where “$\pm$” corresponds to $\nu\geq 1$ and $\nu \leq 1$, respectively. The results are depicted in the right panel of Fig. \ref{ts(nu)}. Remarkably, for each case, a prominent discontinuity in the slope emerges around the critical point. Since such phenomenon was never found in the literature as far as we know, we shall explore it further. 

By evaluating the derivative $\fft{dt_*}{dz_*}$ near the critical point, we find that $\lim\limits_{z_*\rightarrow1^+}\fft{dt_*}{dz_*}\neq\lim\limits_{z_*\rightarrow1^-}\fft{dt_*}{dz_*}$ for any given $\nu\neq 1$. To see this, we expand the derivative near the critical point. We set $\hat{z_s}=z_s-1$ and $\hat{z_l}=1-z_l$. For $\nu>1$, we find
\bea\label{nularge}
\fft{dt_*}{dz_*}\Big|_{\nu>1}&=&\fft{1}{\sqrt{3}}-\fft{\sqrt{3}-1}{3}\hat{z_s}+O(\hat{z_s}^2)\nn\\
&=&-\fft{1}{\sqrt{3}}-\fft{\sqrt{3}+1}{3}\hat{z_l}+O(\hat{z_l}^2)\,,
\eea
whilst for $\nu<1$
\bea\label{nusmall}
\fft{dt_*}{dz_*}\Big|_{\nu<1}&=&-\fft{1}{\sqrt{3}}+\fft{\sqrt{3}+1}{3}\hat{z_s}+O(\hat{z_s}^2)\nn\\
&=&\fft{1}{\sqrt{3}}+\fft{\sqrt{3}-1}{3}\hat{z_l}+O(\hat{z_l}^2) \,.
\eea
 Technically such discontinuity arises because of the square root in (\ref{t*(z)}) . The physical explanation is the transition between the two phases occurs at a given temperature $T>T_c$ for $\nu>1$ and $T<T_c$ for $\nu<1$. This is the case on the $t-z$ plane. However, one may wonder why the discontinuity disappears on the $t-\nu$ plane. The reason is for the two branch solutions $t_*-1\propto |z_*-1|/\sqrt{3}\,,\nu-1\propto \sqrt{3}\,|z_*-1|$ for $\nu>1$ whilst $t_*-1\propto -|z_*-1|/\sqrt{3}\,,\nu-1\propto -\sqrt{3}\,|z_*-1|$ for $\nu<1$. For both cases, the derivative $\fft{\partial t_*}{\partial \nu}$ near the critical point equals $\fft{1}{3}$ on both two sides of the critical point.

Last but not least, in the limit $\nu\rightarrow 0$, our coexistence black holes obey $z_s\rightarrow\infty\,,z_l\rightarrow 0$, according to (\ref{zslnu}). This means the transition occurs between a zero size black hole and an infinitely large black hole. This is of course unphysical and matches with the fact that no transition occurs for classical BTZ black holes having a finite size.

\section{Critical exponents and violation of scaling law}

We move to compute the various critical exponents for $U-\nu$ criticality. The exponents $\alpha$, $\beta$, $\gamma$, $\delta$ are defined as follows:\\
$\bullet$ The exponent $\alpha$ describes the behavior of the specific heat at constant chemical potential $u$
\bea
C_u=\left(\fft{\partial s}{\partial t}\right)_u\propto|t-1|^{-\alpha}
\eea
$\bullet$ The exponent $\beta$ describes the behavior of the order parameter $\eta=u_l-u_s$ (the difference between the chemical potential of the large black hole phase $u_l$ and the small black hole phase $u_s$) on a given isotherm
\bea
\eta=u_l-u_s\propto|t-1|^{\beta}
\eea
$\bullet$ The exponent $\gamma$ determines the behavior of the isothermal compressibility $\kappa_T$
\bea
\kappa_T=-\left(\fft{\partial u}{\partial \nu}\right)_t\propto|t-1|^{-\gamma}
\eea
$\bullet$ The exponent $\delta$ is defined on the critical isotherm $t=1$ as:
\bea
\nu-\nu_c\propto|u-u_c|^{\delta}
\eea
To compute these exponents, we need extract the behaviors of various physical quantities near the critical point. We set $\hat{t}=t-1$, $\hat{z}=z-1$, $\hat{\nu}=\nu-1$ and $\hat{u}=u-1$. 

Firstly, at constant $u$, one has
\bea
0=du=\fft{\partial u}{\partial z}dz+\fft{\partial u}{\partial \nu}d\nu=0 \quad\Rightarrow\quad \left.\fft{d\nu}{dz}\right|_u=-\left(\fft{\partial u}{\partial z}\right)\left/\left(\fft{\partial u}{\partial \nu}\right)\right. \,,
\eea
which gives rise to
\bea
\left(\fft{\partial s}{\partial t}\right)_u=\fft{\fft{\partial s}{\partial z}+\fft{\partial s}{\partial \nu}\left.\fft{d\nu}{dz}\right|_u}{\fft{\partial t}{\partial z}+\fft{\partial t}{\partial \nu}\left.\fft{d\nu}{dz}\right|_u}\,.
\eea
Expanding it near the critical point, we find
\bea
C_u=\fft{17}{4}+O(\hat{t})\,.
\eea
This implies $\alpha=0$. Secondly, according to (\ref{nularge}) and (\ref{nusmall}) and the discussions below
\bea
t&=&1\pm\fft{1}{\sqrt{3}}\hat{z}+O(\hat{z}^2)\nn\\
&=&1+\fft{1}{3}\hat{\nu}+O(\hat{\nu}^2)\,,
\eea
where the “$\pm$” corresponds to the small/large (large/small) black hole side for $\nu>1$ ($\nu<1$), respectively.  This gives $t-1\propto|z-1|\propto |\nu-1|$, which will help us extract various critical exponents.

Expanding $u$ near the critical point by using (\ref{small solution}), we deduce
\bea\label{critu}
u=1+\fft{4\pm5\sqrt{3}}{6}\hat{z}+O(\hat{z}^2)\,,
\eea
where again “$\pm$” corresponds to the small/large (large/small) black hole side for $\nu>1$ ($\nu<1$). This implies to leading order 
\be\label{uhat} \hat{u}=\fft{15\pm 4\sqrt{3}}{6}\,\hat{t}\quad \Rightarrow \quad u_l-u_s=-\fft{4\sqrt{3}}{3}|t-1| \,,\ee
valid to both $\nu>1$ and $\nu<1$. This leads to $\beta=1$.

Besides, on a given isothermal
\bea\label{dzdnut}
0=dt=\fft{\partial t}{\partial z}dz+\fft{\partial t}{\partial \nu}d\nu \quad\Rightarrow\quad \left.\fft{dz}{d\nu}\right|_t=-\left(\fft{\partial t}{\partial \nu}\right)\left/\left(\fft{\partial t}{\partial z}\right)\right.\,,
\eea
so that
\bea
\left.\fft{\partial u}{\partial \nu}\right|_t=\fft{\partial u}{\partial z}\left.\fft{dz}{d\nu}\right|_t+\fft{\partial u}{\partial \nu}\Big|_z\,.
\eea
Then evaluation of the isothermal compressibility near the critical point yields
\bea\label{exp kappaT}
\kappa_T=\fft{1}{9\hat{t}^2}+O(\hat{t}^{-1})\,,
\eea
which implies $\gamma=2$. This exponent describes divergence of the specific heat at constant $\nu$ at the critical point because of $C_\nu-C_u\propto \kappa_T\propto |\hat{t}|^{-2}$.

Finally, to compute the exponent $\delta$, we perform Taylor expansion for the law of corresponding state $\nu=\nu(t,u)$ around the critical point. We find 
\be \nu=1+3\hat{t}+3\hat{t}^2-\fft{3}{64}\big(72\hat{u}^3-540\hat{t}\hat{u}^2+1254\hat{t}^2 \hat{u}-949\hat{t}^3 \big)+\cdots \,. \ee
Here it should be emphasized that the absence of $\hat{t}\hat{u}$ term in the expansion is of paramount importance for critical behavior of various physical quantities (when this term appears it will inevitably give rise to the mean field theory results). In fact, using the expansion, it is straightforward to show that $\beta=1$ and $\gamma=2$ according to the Maxwell's area law, consistent with our previous derivations.  Clearly on the critical isotherm $\hat{t}=0$, 
\be \nu-1=-\fft{27}{8}(u-1)^3 \quad\Rightarrow\quad \delta=3 \,.\ee
This completes our derivations. 

It is worth emphasizing that the exponents $\alpha=0\,,\beta=1\,,\gamma=2\,,\delta=3$ are significantly different from the mean field theory results $\alpha=0\,,\beta=1/2\,,\gamma=1\,,\delta=3$. This can be attributed to the quantum backreactions of conformal fields in AdS$_3$. We point out the same set of exponents was previously obtained for the quantum anomaly corrected black holes as well \cite{Hu:2024ldp}. 

However, it is quite surprising that this set of exponents violates  the scaling law
\be 2-\alpha=\beta(1+\delta) \,,\label{scalinglaw1}\ee
but still obeys the one
\be \gamma=\beta(\delta-1)\,.\label{scalinglaw2}\ee
This is also pointed out in \cite{Hu:2024ldp} but a proper interpretation is lacking. We recall that the above scaling laws follow from the universal behavior of (the singular part) of thermodynamic potential $f_s$ near the critical point. The two-scale factor universality hypothesis \cite{stauffer,widom,stanley,peskin} suggests that 
\be f_s=c_1 \hat{t}^{\beta(1+\delta)}y(c_2\, m\,\hat{t}^{-\beta}) \,,\label{twoscale}\ee
where $m$ is the order parameter and $y(x)$ is a smooth function obeying suitable conditions (which are however unimportant in our discussions). The constants $c_1\,,c_2$ are two scale factors, depending on material and differentiating between systems even in the same universality class. Since absolute value of these factors is unimportant, we will not write them down explicitly in the following.

Consider for example the transition occurs below the critical temperature. Then the value of the order parameter is found by minimizing $f_s$ with respect to $m$. In the scaling region, this occurs at the minimum $x_0$ of the function $y(x)$, implying $m\propto \hat{t}^\beta$. 

To compute the specific heat $C_m$, we differentiate $f_s$ twice with respect to temperature 
\be C_m\propto \hat{t}^{\beta(1+\delta)-2}\equiv \hat{t}^{-\alpha} \qquad\Rightarrow\qquad 2-\alpha=\beta(1+\delta) \,.\ee
This gives rise to the scaling law (\ref{scalinglaw1}).

The external field $h$ conjugate to the order parameter $m$ can be evaluated to be 
\be h=-\fft{\partial f_s}{\partial m}=\hat{t}^{\beta\delta}\,y'(m \,\hat{t}^{-\beta}) \,.\ee
In the scaling region $m\propto \hat{t}^\beta$ and hence $h\propto m^\delta$. The inverse of the above relation gives
\be m=\hat{t}^\beta w(h\, \hat{t}^{-\beta\delta}) \,.\ee
By definition, the isothermal compressibility can be evaluated as
\be \kappa_T=\Big( \fft{\partial m}{\partial h}\Big)_t\propto \hat{t}^{-\beta(\delta-1)}\equiv \hat{t}^{-\gamma}\qquad\Rightarrow \qquad \gamma=\beta(\delta-1)  \,,\ee
which leads to the scaling law (\ref{scalinglaw2}). 

From above discussions, it is easy to see that emergence of these scaling laws should be consistent with the scaling behavior of the external field $h$ and the order parameter $m$. As a consequence, while changing the power of $\hat{t}$ in the two-scale factor hypothesis can modify the scaling laws, it will also predict incorrect behavior for either the external field or the order parameter (or both of them).  Therefore, to interprete the violation of the scaling law (\ref{scalinglaw1}) in a self-consistent manner, we have to develop a three-scale factor hypothesis for the thermodynamic potential
\be f_s=c_0\,\hat{t}^{2-\alpha}+c_1 \hat{t}^{\beta(1+\delta)}y(c_2\, m\,\hat{t}^{-\beta}) \,,\ee
where $c_0$ is an extra scale factor. The presence of this term can be viewed as the thermodynamic potential of the normal (or symmetrical) phase in which $m=0$. The separation of the exponent $\alpha$ and the order parameter plays a pivotal role in our proposal. Clearly our assumption is still consistent with the scaling behavior of $h$ and $m$ as well as the scaling law (\ref{scalinglaw2}). Assume $2-\alpha<\beta(1+\delta)$ and differentiating $f_s$ twice with respect to temperature yields to leading order
\be C_m\propto c_0\,\hat{t}^{-\alpha} \,.\ee
This predicts the correct scaling behavior of the specific heat and generally evades the constraint from the scaling law (\ref{scalinglaw1}). Physically it tells us that the violation of (\ref{scalinglaw1}) means: i) the leading term of thermodynamic potential determines the singular behavior of the specific heat $C_m$; ii) but the critical behavior of the order parameter $m$ is determined by the subleading term. We do not have any idea why this could happen. Yet it is easy to see that whenever the scaling of order parameter is determined by the leading term of the thermodymamic potential, the scaling law (\ref{scalinglaw1}) will be satisfied inevitably. In this case, the scale factor $c_0$ will be redundant and our hypothesis simply reduces to the ordinary two-scale factor case. This implies that a general constraint on the exponent $\alpha$ might be
\be  2-\alpha\leq \beta(1+\delta) \,.\label{generalconstaint}\ee 
However, we are aware of that our hypothesis just provides a formal explanation for violation of the scaling law (\ref{scalinglaw1}). It does not tell us when and how this could happen in physics. It will be very interesting to study it further in a renormalization group analysis for certain quantum field theories. We leave this to future research.


\section{With angular momenta corrections}

Now we move to investigate the correction effect of a small angular momenta $J$ on the $U-\nu$ criticality. 
Firstly, to derive the critical point, we adopt the inflection point condition 
\be \Big(\fft{\partial T}{\partial S}\Big)_{\nu\,,J}=\Big(\fft{\partial^2T}{\partial S^2}\Big)_{\nu\,,J}=0 \,.\ee
The resulting expressions are quite lengthy and are not instructive to be presented. Clearly the critical point $\nu_c=z_c=1$ for the static qBTZ black holes is still a solution to the equations, corresponding to the trivial case $J=0$. To search nontrivial solutions, we consider small $J$ corrections
\bea
&&z_c=1+z_2J^2+O(J^3)  \,,\nn\\
&&\nu_c=1+\nu_1J+O(J^2) \,.
\eea
Notice that $z_1$ vanishes identically owing to expansion of the derivatives (\ref{derexpansion}). By using expansion of the rotation parameter $\alpha$
\bea\label{alpha=J}
\alpha=\fft{z+3z^2+2z^3\nu}{2(1+z^2)\sqrt{(1+\nu^2)(1+z\nu)(1-z^3\nu)}}\fft{G_3}{l_3}J+O(J^3)\,.
\eea
 and expanding the derivatives
\bea\label{derexpansion}
&&0=\Big(\fft{\partial T}{\partial S}\Big)_{\nu\,,J}=\Big(\fft{\partial T}{\partial S}\Big)_{\nu\,,J}^{(0)}+\Big(\fft{\partial T}{\partial S}\Big)_{\nu\,,J}^{(2)}J^2+O(J^3)  \,, \nn\\
&&0=\Big(\fft{\partial^2T}{\partial S^2}\Big)_{\nu\,,J}=\Big(\fft{\partial^2T}{\partial S^2}\Big)_{\nu\,,J}^{(0)}+\Big(\fft{\partial^2T}{\partial S^2}\Big)_{\nu\,,J}^{(2)}J^2+O(J^3)\,,
\eea
we are able to obtain new sets of critical points receiving angular momenta corrections by solving the equations order by order. However,  the leading corrections to critical points are unphysical
\bea
\left(\fft{\partial T}{\partial S}\right)_{\nu\,,J}^{(2)}=\fft{G_3\left(1944G_3^2+\nu_1^2 l_3^2\right)}{3\sqrt{2}\pi^2 l_3^4}=0\quad &\Rightarrow& \quad \nu_1=\pm\fft{18\sqrt{6}G_3}{l_3}i \,,\nn\\
\left(\fft{\partial^2T}{\partial S^2}\right)_{\nu\,,J}^{(2)}=\fft{9G_3^2\left(4 l_3^2 z_2-1701G_3^2\right)}{2\pi^3 l_3^5}=0\quad &\Rightarrow& \quad z_2=\fft{1701G_3^2}{4l_3^2}\,.
\eea
In fact, we also find $z_3=\pm\fft{2430\sqrt{6}G_3^3}{l_3^3}i$ according to the third-order term. These imaginary values of the correction coefficients imply that there does not exist any nontrivial critical point for a given angular momenta.
\begin{figure}
  \centering
  \includegraphics[width=210pt]{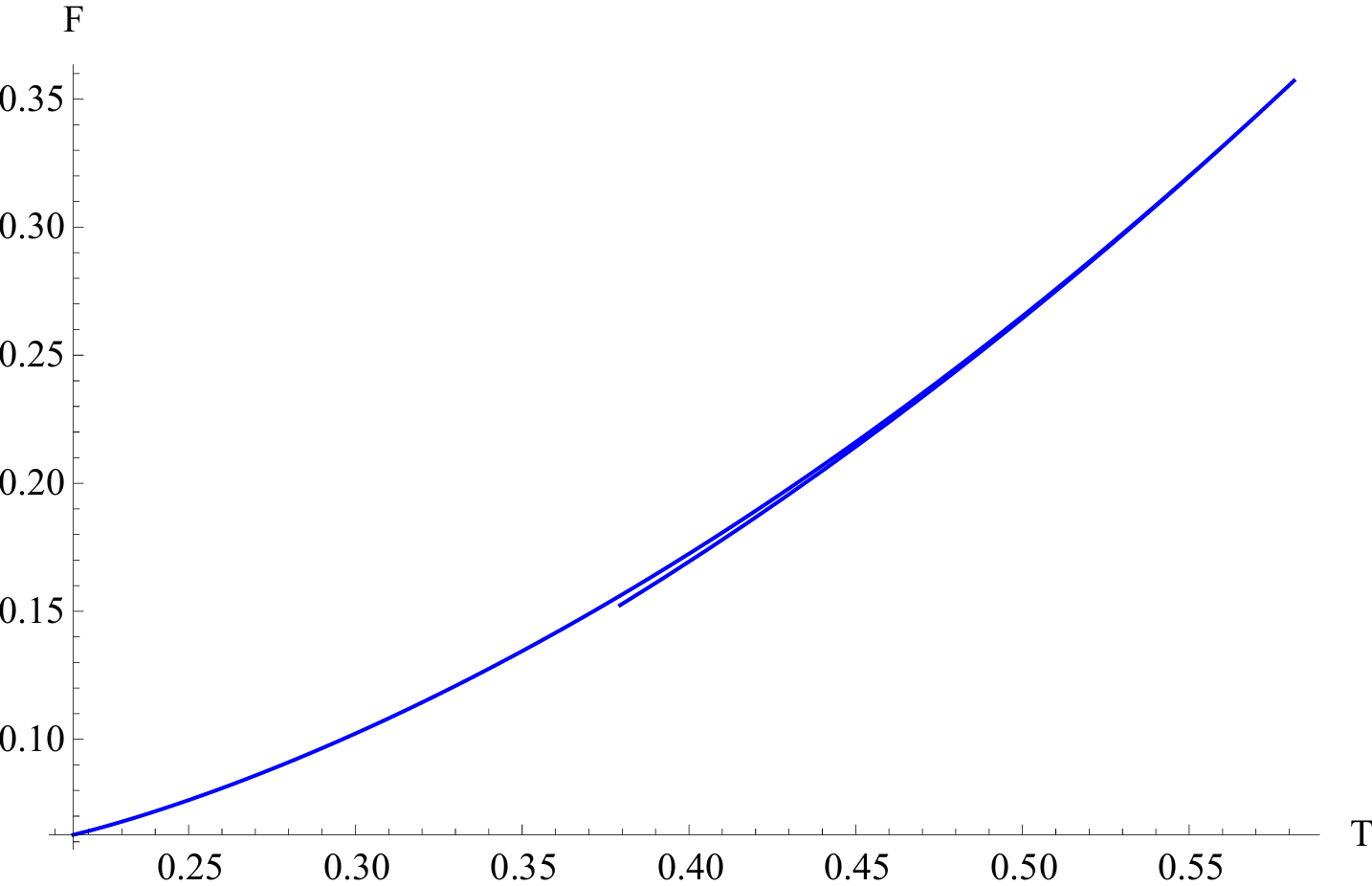}
  \includegraphics[width=210pt]{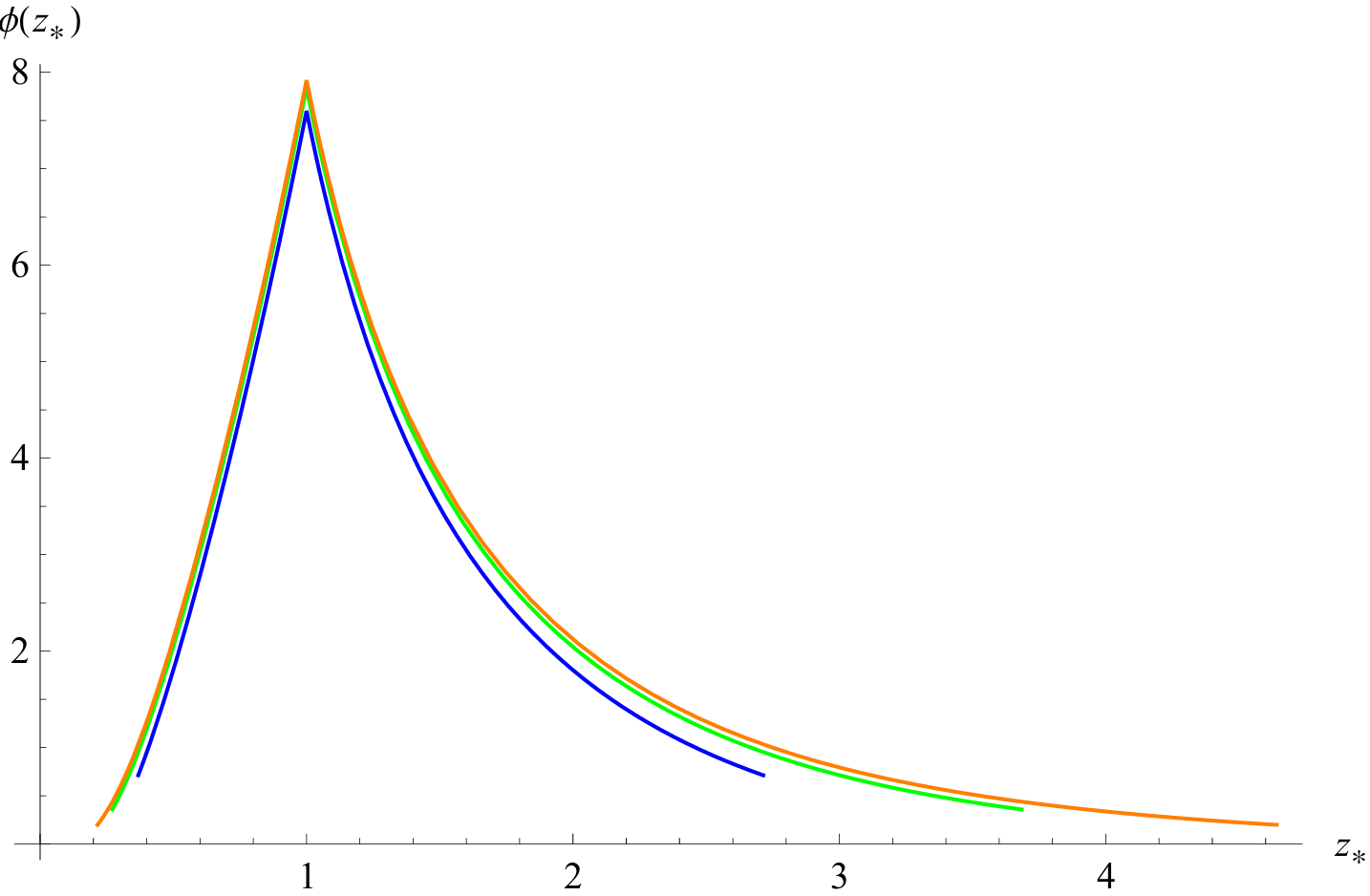}
  \caption{Left: the behavior of free energy for a rotating qBTZ black hole with $\nu=1/100$. Right: the function of $\phi(z_*)$ for $\nu=1/20$ (blue), $\nu=1/50$ (green) and $\nu=1/100$ (orange).  }
\label{ftdel}\end{figure}

The remaining issue is whether we can interpret the static critical point as $J_c=0$ so that there exists phases transition for a given nonzero angular momenta? Intuitively the answer is yes since a small $J$ correction does not alter the equation of state and the characteristic behavior of free energy very much. While  this is formally correct, the problem is a nonzero angular momenta will strongly constrain the size of qBTZ black holes (recall $z\leq \nu^{-1/3}$ according to (\ref{rotating quantities}) or (\ref{alpha=J})). The consequence is only a single stable black hole phase can exist and no transitions will occur.

To show this explicitly, consider an arbitrarily small $J$ so that existence of phases transition is essentially determined by the condition (\ref{small solution}) in the static case. Since $z\leq \nu^{-1/3}$ and $z_sz_l=1$, the black holes size will be constrained to be in the region 
\be \nu^{1/3}\leq z_* \leq \nu^{-1/3}\quad \mm{and}\quad \nu\leq 1 \,.\label{regime}\ee 
The behavior of free energy for a rotating qBTZ black hole obeying this constraint is depicted in the left panel of Fig. \ref{ftdel}. It is immediately seen that the characteristic swallow tail behavior disappears for various values of $\nu$. We shall show that this is always true for all $\nu<1$.  

Consider the function
\be \phi(z_*)\equiv (z_*+1/z_*)^3-\sqrt{(z_*+1/z_*)^6-64}-8\nu \,.\ee
If two stable black hole phases coexist in the region (\ref{regime}), the function $\phi$ should have two positive real roots. However, this cannot happen. As depicted in the right panel of Fig. \ref{ftdel}, the function $\phi(z_*)$ increases (decreases) as $z_*$ increases for the large (small) size black holes. Therefore the function $\phi(z_*)$ takes the maximum at the critical point $z_*=1$ and the minimum $\phi_{\mm{min}}$ occurs at $z_*=\nu^{\pm1/3}$. However, the minimal value $\phi_{\mm{min}}$ is positive definite and decreases as $\nu$ approaches to zero. In the small $\nu$ limit, $\phi_{\mm{min}}\approx 24\nu$ is always positive definite. Thus, there exists only a single stable black hole phase and no transition will occur given an arbitrarily small angular momenta.

\section*{Acknowledgments}

Z.Y. Fan was supported in part by the National Natural Science Foundations of China with Grant No. 11873025.

\appendix

\end{document}